\documentclass[a4paper,fleqn,usenatbib]{mnras}   %mn2e} %{mnras}
\usepackage{xcolor}
\usepackage[T1]{fontenc}
\usepackage{ae,aecompl}
\usepackage{graphicx}	% Including figure files
\usepackage{amsmath}	% Advanced maths commands
\usepackage{amssymb}	% Extra maths symbols
\usepackage{soul}
\newcommand{\feiii}{\textrm{Fe}\textsc{iii}}
\newcommand{\feii}{\textrm{Fe}\textsc{ii}}
\newcommand{\carbonio}{\textrm{C}}
\newcommand{\ferro}{\textrm{Fe}}
\newcommand{\silicio}{\textrm{Si}}
\newcommand{\azoto}{\textrm{N}}

\newcommand{\hi}{\textrm{H}\textsc{i}}
\newcommand{\heialone}{\textrm{He}\textsc{i}}
\newcommand{\heiialone}{\textrm{He}\textsc{ii}}

\newcommand{\oii}{[\textrm{O}\textsc{ii}]}

\newcommand{\mgii}{\textrm{Mg}\textsc{ii}~\ensuremath{\lambda2800}}

\newcommand{\oialone}{\textrm{O}\textsc{i}}
\newcommand{\oialonesq}{[\textrm{O}\textsc{i}]}

\newcommand{\caiialonesq}{[\textrm{Ca}\textsc{ii}]}
 
\newcommand{\ha}{\ifmmode {\rm H}\alpha \else H$\alpha$\fi}
\newcommand{\hb}{\ifmmode {\rm H}\beta \else H$\beta$\fi}
\newcommand{\lya}{\ifmmode {\rm Ly}\alpha \else Ly$\alpha$\fi}
\newcommand{\pg}{\ifmmode {\rm P}\gamma \else Pa$\gamma$\fi}
\newcommand{\lyb}{\ifmmode {\rm Ly}\beta \else Ly$\beta$\fi}
\newcommand{\lyg}{\ifmmode {\rm Ly}\gamma \else Ly$\gamma$\fi}

\newcommand{\ciiidoub}{\textrm{C}\textsc{iii}]\ensuremath{\lambda\lambda1907,1909}}

\newcommand{\siiiiblue}{\textrm{Si}\textsc{iii}]\ensuremath{\lambda1883}}
\newcommand{\siiiired}{\textrm{Si}\textsc{iii}]\ensuremath{\lambda1892}}
\newcommand{\siiiidoub}{\textrm{Si}\textsc{iii}]\ensuremath{\lambda\lambda1883,1892}}

\newcommand{\neiii}{\textrm{Ne}\textsc{iii}]\ensuremath{\lambda3869}}

\newcommand{\heii}{\textrm{He}\textsc{ii}\ensuremath{\lambda1640}}

\newcommand{\flyc}{\ifmmode  \mathrm{f}_\mathrm{esc}\mathrm{(LyC)} \else $\mathrm{f}_\mathrm{esc}\mathrm{(LyC)}$\fi}

\def\kms{km s$^{-1}$}

\def\ergs{\ifmmode \mathrm{erg\hspace{1mm}s}^{-1} \else erg s$^{-1}$\fi}

\def\micron{\ifmmode \mu\mathrm{m} \else $\mu$m\fi}
\def\msun{\ifmmode \mathrm{M}_{\odot} \else M$_{\odot}$\fi}
\def\msunyr{\ifmmode \mathrm{M}_{\odot} \hspace{1mm}{\rm yr}^{-1} \else $\mathrm{M}_{\odot}$ yr$^{-1}$\fi}
\def\zsun{\ifmmode Z_{\odot} \else Z$_{\odot}$\fi}
\def\lsun{\ifmmode L_{\odot} \else L$_{\odot}$\fi}
\def\mstar{\ifmmode \mathrm{M}_{\star} \else M$_{\star}$\fi}

\title[\lya\ Fluorescence emission at z=2.37]
{Probing the circum-stellar medium 2.8 Gyr after the Big Bang:
detection of Bowen fluorescence in the Sunburst arc}
\author[E.~Vanzella et al.]{
\parbox[t]{\textwidth}{E.~Vanzella$^1$\thanks{E-mail: eros.vanzella@inaf.it}\thanks{Based on observations collected at the European Southern Observatory for Astronomical
research in the Southern Hemisphere under ESO programmes 0103.A-0688, 60.A-9507, 297.A-5012.}, M. Meneghetti$^1$, A. Pastorello$^2$, F. Calura$^1$, E. Sani$^3$, G. Cupani$^4$, G.B. Caminha$^5$, M. Castellano$^6$, P. Rosati$^{7,1}$, V. D'Odorico$^4$, S. Cristiani$^4$, C. Grillo$^8$, A. Mercurio$^9$, M. Nonino$^4$, G.B.~Brammer$^{10}$ and H.~Hartman$^{11}$
}
\vspace*{8pt}\\
$^1$INAF -- Osservatorio di Astrofisica e Scienza dello Spazio, via Gobetti 93/3, 40129 Bologna, Italy\\
$^{2}$INAF -- Osservatorio Astronomico di Padova, Vicolo Osservatorio 5, 35122, Padova, Italy\\
$^3$European Southern Observatory, Alonso de Cordova 3107, Casilla 19, Santiago 19001, Chile \\
$^4$INAF -- Osservatorio Astronomico di Trieste, via G. B. Tiepolo 11, I-34143, Trieste, Italy\\
$^5$Kapteyn Astronomical Institute, University of Groningen, Postbus 800, 9700 AV Groningen, The Netherlands\\
$^6$INAF -- Osservatorio Astronomico di Roma, Via Frascati 33, I-00078 Monte Porzio Catone (RM), Italy\\
$^7$Dipartimento di Fisica e Scienze della Terra, Universit\`a degli Studi di Ferrara, via Saragat 1, I-44122 Ferrara, Italy\\
$^8$Dipartimento di Fisica, Universit\`a  degli Studi di Milano, via Celoria 16, I-20133 Milano, Italy\\
$^9$INAF -- Osservatorio Astronomico di Capodimonte, Via Moiariello 16, I-80131 Napoli, Italy\\
$^{10}$Cosmic Dawn Center, Niels Bohr Institute, University of Copenhagen, Juliane Maries Vej 30, DK-2100 Copenhagen \O, Denmark\\
$^{11}$Department of Materials Science and Applied Mathematics, Malm\"o University, SE-20506 MALM\"O, Sweden\\
}

\begin{document}
\date{}
\maketitle

\begin{abstract}
We discovered Bowen emission arising from a strongly lensed (i.e., with magnification factor $\mu>20$) source hosted in the Sunburst arc at $z=2.37$. We claim this source is plausibly a transient stellar object and study the unique ultraviolet lines emerging from it. 
In particular, narrow ($\sigma_{v} \simeq 40$ \kms) ionisation lines of \ferro\ fluoresce after being exposed to \lya\ radiation \heialone\ and \heiialone\ 
that pumps selectively their atomic levels. 
Data from VLT/MUSE, X-Shooter and ESPRESSO observations (the latter placed at the focus of the four UTs) at increasing spectral resolution of R=2500, 11400 and R=70000, respectively,
confirm such fluorescent lines are present since at least 3.3 years ($\simeq 1$ year rest-frame). 
Additional \ferro\ forbidden lines have been detected, while \carbonio\  and \silicio\  doublets probe an electron density $n_e \gtrsim 10^{6}$~cm$^{-3}$. Similarities with the
spectral features observed in the circum-stellar Weigelt blobs of Eta~Carinae probing the circum-stellar dense gas condensations in radiation-rich conditions are observed.
We discuss the physical origin of the transient event, which remains unclear. We expect such transient events (including also supernova or impostors) will be easily recognised with ELTs thanks to high angular resolution provided by adaptive optics and large collecting area, especially in modest ($\mu < 3$) magnification regime. 
\end{abstract}


\begin{figure*}
\centering
\includegraphics[width=12cm]{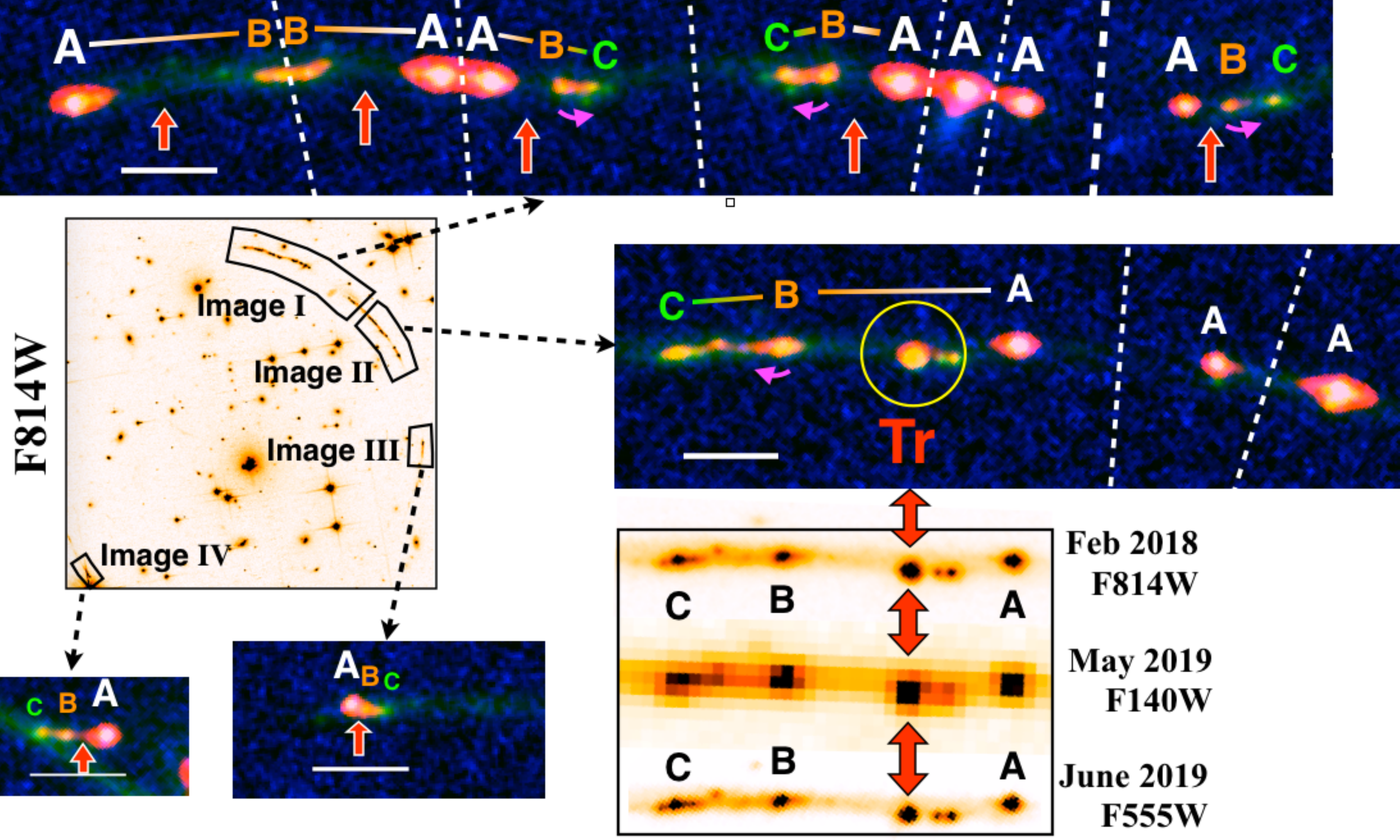}
\caption{The four multiple arcs I, II, III and IV are shown with RGB color images (red=F160W, green=F814W and blue=F606W), in which knots A are marked, as unambiguously identified by \citet{rivera19}. Knots {\tt B} and {\tt C} are also identified following the expected parity introduced by lensing. The object {\tt Tr} is indicated with a yellow circle and its expected multiple images located between knots {\tt A} and {\tt B} are marked with red arrows. A non-transient object would appear 8 times and as bright as knot {\tt A}. Magenta small arrows mark a feature emerging from the red band, F160W, only on one side of knot {\tt B}. Such a tail replicates accordingly to multiple images (presence) and parity (orientation). The bottom-right panel shows the location of {\tt Tr} split over multi-epoch HST observations, from Feb. 2018 to June 2019.}
\label{pano_HST}
\end{figure*}

\section{Introduction}

Three main excitation mechanisms are involved in the
explanation of the occurrence of bright emission lines in astrophysical sources: 
photoionization followed by recombination, photoexcitation to an upper level and collisional excitation. 
Among the processes belonging to the second category is the Bowen fluorescence mechanism \citep[][]{bowen34}, characterised by photoexcitation
by accidental resonance (PAR), i.e. the coincidence
between the wavelength of a strong emission line and a transition between different atomic levels of another element.

A known example is the emission line of neutral oxygen,
\oialone, at 8446\AA, which can be caused by a wavelength coincidence
between the brights \hi~\lyb\ line (1025.72\AA) and the absorption line of \oialone\ at 1025.77\AA, followed by a successive cascade \citep[][]{bowen47}.
The Bowen fluorescence involving other elements like \carbonio, \silicio, \ferro, \azoto\ has also been observed in
various  and  diverse  astrophysical  systems such  as  planetary  nebulae \citep[e.g.,][]{wey69}, Be stars \citep[][]{merrill56}, X-ray binary stars \citep[e.g.][]{clinto75}, Wolf-Rayet stars \citep[e.g.,][]{crowther07}, 
symbiotic stars \citep[][]{walle91}, local Seyfert 1 nuclei and quasars \citep[e.g.,][]{marziani14,netzer85}, and nearby supernovae \citep{pastore02,gra17,lelo19}. 
In particular, a plethora of fluorescing emission lines in the visible and near-IR have been observed emerging from compact gas condensations located near the Luminous Blue Variable (LBV) star of $\eta$~Carinae (or $\eta$~Car hereafter).
Specifically, the massive and luminous star in $\eta$~Car ($\gtrsim 100$\msun\ and $L \gtrsim 6 \times 10^{6} L_{\odot}$) is expelling an enormous amounts of material into its surrounding, where three compact gas condensations ($N_{H} \simeq 10^{5-10}$ cm$^{-3}$, depending
on the ionic diagnostics used, \citealt{H99}) at a distance of $10^{16}$ cm from the star (a few light days) have been  studied in great detail. Known as the ``Weigelt blobs'' \citep{w86}, such condensations are slowly moving ($\sim 50$ \kms) and producing many hundreds intense, narrow fluorescing emission lines including lasing emission\footnote{Lasing emission is dominated by stimulated rather than spontaneous emission.}, unlike the spectrum of any other known object \citep{joh07}.
Is it possible to observe such spectral features at cosmological distance and detect gas condensations in the vicinity of a remote massive star? 
Strongly lensed supernova events have been recently detected at cosmological distances \citep[with the most spectacular case dubbed Refsdal, at z=1.49,][]{kelly15,grillo16}.
In this Letter we report on the {\it first} detection of Bowen (fluorescent) emission from a strongly lensed candidate transient stellar object at cosmological distance (z=2.37, corresponding to 2.8 Gyr after the Big-Bang) probing circum-stellar gas condensation, 
hosted in a giant gravitational arc known as {\it Sunburst} arc \citep[e.g.,][]{rivera19}.
The transient stellar object we report here is highly magnified (with a magnification factor $\mu>20$), implying that sub-regions of the arc are probed down to a few tens of parsec \citep[][]{vanz20}.
We assume a flat cosmology with $\Omega_{M}$= 0.3,
$\Omega_{\Lambda}$= 0.7 and $H_{0} = 70$ km s$^{-1}$ Mpc$^{-1}$.

\section{The transient in the Sunburst arc}
An exceptionally bright (mag $\simeq 18$) gravitationally lensed multiple imaged arc at z=2.37 was discovered by \citet{dahle16}, in which several star-forming regions down to a few tens of parsec scale have been recognised. 
Based on the unique signature provided by the ionization leakage, \citet{rivera19} unambiguously identified one knot replicated 12 times due to strong gravitational lensing.
\citet{vanz20} studied in detail such a knot (dubbed {\tt A} in Figure~\ref{pano_HST}) and identified other nearby additional features following the parity introduced by lensing. In particular, knot {\tt A} is part of three major star-forming regions, shown in Figure~\ref{pano_HST} as {\tt A}, {\tt B} and {\tt C}.
It is worth noting that the identification of knots {\tt B} and {\tt C} is facilitated by the unambiguous 12 detections of knot {\tt A}. Moreover, Figure~\ref{pano_HST} also shows multiple images of a signature emerging in the F160W band (indicated with magenta arrows and following the expected parity), further confirming the identification of knot {\tt B}. 
Consequently, also knot {\tt C} fits into the global picture, following the multiplicity and parity
(see Figure~\ref{pano_HST}, or appendix A of \citealt{vanz20}). Unexpectedly, {\tt Tr} appears only in the arc II, in between knots {\tt B} and {\tt A} (Figure~\ref{pano_HST}). {\tt Tr} is spatially unresolved and has a magnitude comparable to {\tt B} and {\tt A} (F814W $\simeq 22$), implying that it would be easily detectable 8 times, in between knots {\tt A} and {\tt B}. However, it is observed only one time, strongly suggesting {\tt Tr} is a transient object. 
Another relevant quantity is the observed magnitude of {\tt Tr} that lies in the range $-20.3 < M_{2000}< -18.6$, corresponding to two extremes magnifications values, $\mu=20$ and $\mu=100$, respectively. The first value is the minimum magnification derived in \citet{vanz20} (based on geometrical arguments), the second is based on the assumption that {\tt Tr} is not close to a critical line, being located within knots {\tt A} and {\tt B}, where no critical lines are expected. $\mu=100$ is therefore a likely upper limit. 
Finally, as discussed in the next section, peculiar spectral features emerge from {\tt Tr}, that, however, are not observed in any other position of the arcs.

To summarize, if {\tt Tr} was not a transient, we would detect it as bright as knots {\tt A} (or {\tt B}) 8 times, as well as its spectral features, spread over the arcs I, II, III and IV (no significant magnification variation is expected within the triplet {\tt A}, {\tt B}, {\tt C}). Using these arguments, we propose that {\tt Tr} is plausibly a transient. As discussed below, this interpretation is corroborated by some spectral similarity with supernovae or non-terminal transients whose ejecta interact with circumstellar material \citep{pastore02,gra17}.

\begin{figure}
\centering
\includegraphics[width=7.5cm]{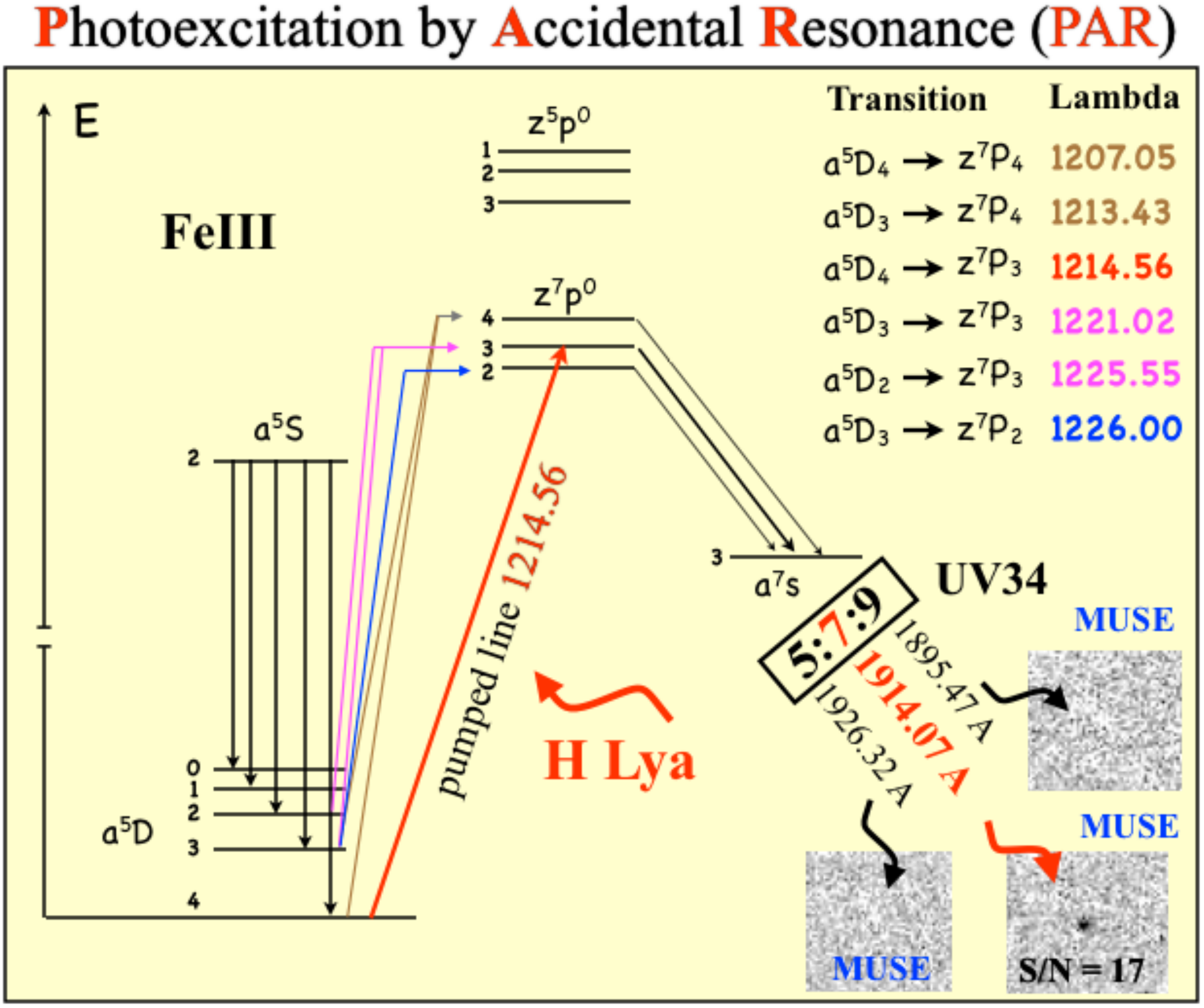}
\caption{The Grotrian diagram showing the PAR emission from the triplet of Fe~{\sc III}~UV34 (adapted from \citealt{joh00}), with indicated the relative ratios among the three lines in the case of not \lya\ pumping (5:7:9, see text). On the top-right the wavelengths of the transitions are reported (also shown with the same color in the diagram). In the bottom-right the three continuum$-$subtracted MUSE slices ($dv = 300$ \kms), showing the three wavelengths of UV34, among which only the 1914.07\AA\ line emerges, with rest-frame equivalent width of $0.56\pm0.05$\AA.}
\label{GRO}
\end{figure}

\section{Spectroscopic observations of Tr}

Figure~\ref{pano_HST} shows that {\tt Tr} was present during HST imaging in February and June 2019 (PI. H. Dahle, ID~15101). However, as discussed below, the peculiar lines associated to such a transient exist since 2016 (from MUSE DDT programme 297.A-5012(A), PI. Aghanim) till September 2019, as our ESPRESSO observations demonstrate, implying it exists since at least 11.9 months rest-frame.

\subsection{VLT/MUSE, X-Shooter and ESPRESSO observations of {\tt Tr}}

VLT/MUSE observations of the Sunburst arc were performed in May-August 2016 and presented in \citet{vanz20}, in which the {\tt Tr} object was identified.
Subsequently, dedicated VLT/X-Shooter observations (Prog. 0103.A-0688, PI Vanzella) of {\tt Tr} have been acquired on 1,2 May $-$ 2,3 August 2019 with R=11400, and finally VLT/ESPRESSO at the focus of the four VLT/UTs was used in early September 2019 at resolution R=70000, as part of the science verification program of the instrument (ID 60.A-9507(A), PI Vanzella).
X-Shooter data were reduced with the pipeline released by ESO \citep{modigliani10}, following the standard approach described, e.g., in \citet{vanz20}. We explicitly defined the windows to extract the object and estimate the background within the slit. For the {\tt Tr} object we used $2.8\pm0.9$ arcsec, with two background windows at $-3.6\pm1.0$ arcsec and $0.8\pm0.5$ arcsec (positions are referred to the centre of the slit). ESPRESSO data were reduced with the Data Reduction Software (DRS) released by ESO \citep{sosnowska15} and analysed with the Data Analysis Software (DAS) specifically developed for ESPRESSO \citep{cupani19}. 
While VLT/MUSE and X-Shooter are well tested instruments, it is worth commenting on the ESPRESSO observations. {\tt Tr} with mag $\simeq 21.5$ (AB) is considered a faint target for the capabilities of the instrument, furthermore it was observed at a relatively high airmass of 1.7. Such conditions made the simultaneous centering and tracking of the target on each UT relatively challenging. Despite that, two 23-minutes scientific exposures were successfully acquired and reduced. The wavelength-calibrated, sky-subtracted, optimally-extracted spectra of the orders produced by the DRS were combined by the DAS into a spectrum of remarkable quality, in which the continuum is detected at S/N $\simeq$ 3 (per resolution element) together with several emission lines at S/N $\simeq 3$-$10$ (see Figure~\ref{pano_spec}). These values are in agreement with the prediction of the instrument exposure time calculator, and correspond to an estimated peak efficiency of 0.08 (including atmospheric transmission and fiber losses).

\subsection{Photoexcitation by Accidental Resonance: \lya\ pumping}
\begin{figure*}
\centering
\includegraphics[width=15cm]{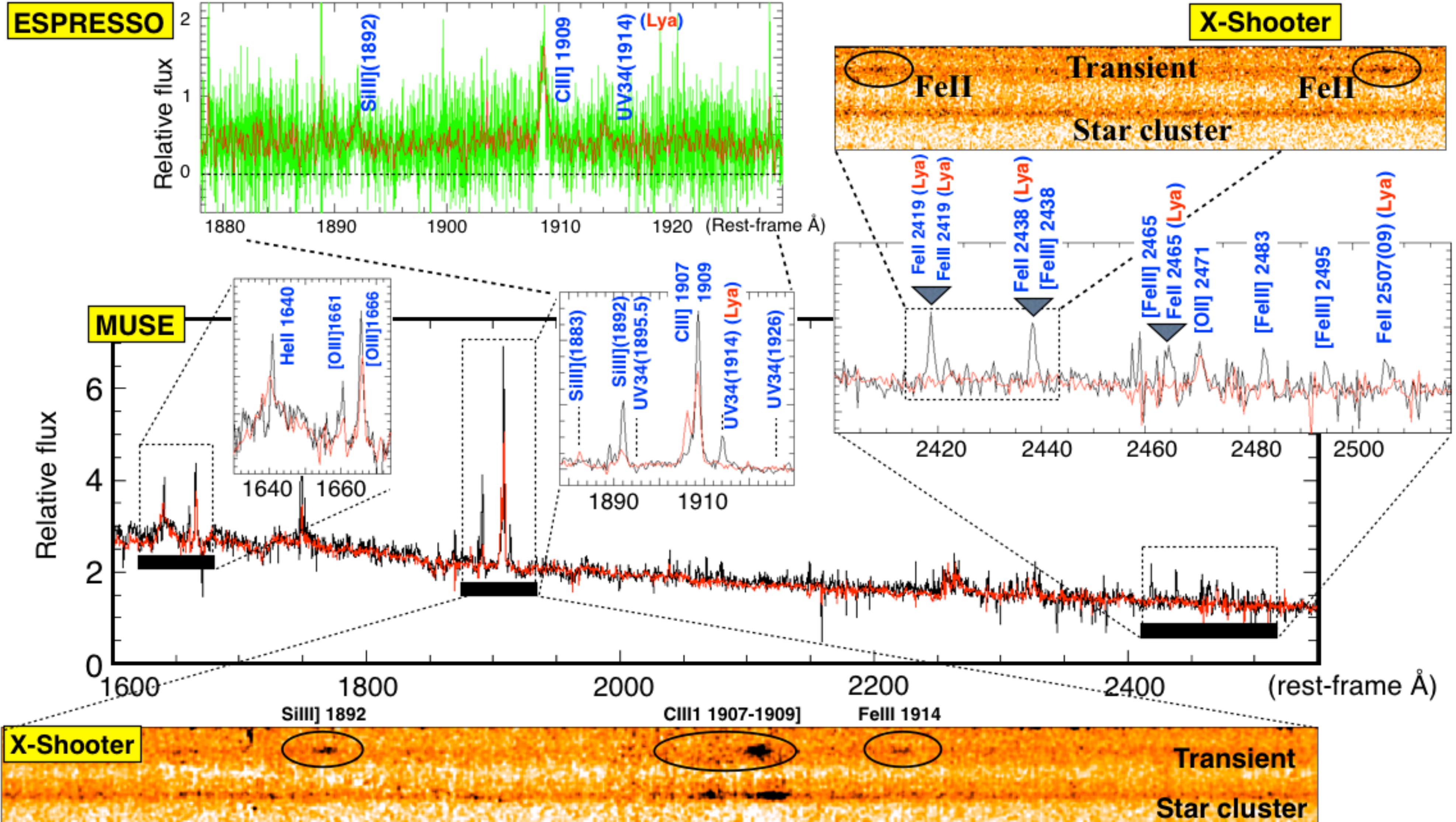}
\caption{A collection of the VLT/MUSE, X-Shooter and ESPRESSO spectroscopic observations in the ultraviolet portion of the spectrum. In the main panel with thick axis, the MUSE spectrum (obtained in June 2016) at R=2500 of the star cluster (knot A, red line) and the transient ({\tt Tr}, black line) are shown, with the insets blowing up the regions around the fluorescence lines 
due to H~\lya\ pumping (see text for details). The two-dimensional X-Shooter data (obtained in May 2019) are also shown with resolution R=11400. The one dimensional spectrum at R=70000 obtained with ESPRESSO during Sept. 2019 at the focus of the 4 VLT/UTs is shown in the top-left (the green/red line corresponds to R=70000/7000). 
The \ferro\ \lya\ pumped lines are marked with a red label.}
\label{pano_spec}
\end{figure*}

 The first spectral feature we noticed is the line emission at 1914\AA~rest-frame (see Figure~\ref{GRO} and \ref{pano_spec}). This line is part of the multiplet UV34, 
 the two lowest excited configurations of \feiii, 3d$^{5}$($^{6}$S)4s~a$^{7}$S $-$ 3d$^{5}$($^{6}$S)4p~z$^{7}$P, which is composed by three lines 1895.473\AA, 1914.066 \AA, and 1926.320\AA\ (Figure~\ref{GRO} shows the Grotrian diagram). As discussed in \citet{joh00} the intensity ratios of the three lines assuming a thermal population 
are 9:7:5 (that is the ratio of the statistical weights of the upper levels).
In the present case, the 1914\AA\ line is detected with S/N=17 and a rest-frame equivalent width of $0.56\pm0.05$\AA, and is more than 10 times stronger than the 1895\AA\ and 1926\AA\ lines (Figures~\ref{GRO} and~\ref{pano_spec}), which suggests a Photoexcitation by Accidental Resonance (PAR) effect is in place, selectively exciting the z$^{7}$P$_{3}$ level. Such emission mechanism has also been detected and deeply investigated by \citet{joh00} for the case of $\eta$~Carinae, in which the same iron line 1914\AA\ is selectively excited by H~\lya\ photons. If the excitation was continuum fluorescence, two lines in UV34 would be observed and in the case of collisional excitation or recombination all three lines would appear.
The UV34 multiplet of \feiii\ can thus be used for determining whether the excitation source is blackbody radiation with substantial flux around 1215\AA\ or substantial spectral compression due to H~\lya\ radiation pumping a single channel.

The presence of additional fluorescence \lya$-$pumped iron lines at 2400-2500\AA\ corroborates such an interpretation.
In particular, \citet{zet12} identify more than 2500 lines emerging from dense gas condensation (Weigelt blobs) in the proximity of the hot star in $\eta$~Car, several of them identified as fluorescent \lya\ pumped transitions. Besides the aforementioned \feiii\ 1914\AA, we find additional \feii\ \lya-pumped lines in \feii\  and \feiii.  Figure~\ref{pano_spec} shows four of them, 2419, 2439, 2464 and 2507-2509\AA, detected in the 2016 MUSE observations (at resolution R=2500), and still present in 2019 X-Shooter observations (at R=11400). These lines are only observed in {\tt Tr}.

Additional forbidden [\feiii] lines are detected at 2483\AA\ and 2495\AA\ rest-frame from the multiplet 3d$^6$ $^5$D$_4$~$-$~3d$^5$($^6$S)4s a$^5$S. Two additional lines of the multiplet are the 2438 and 2465 \AA, blending with the fluorescent lines discussed above. The line ratios of \ciiidoub\AA\ and \siiiidoub\AA\ provide a measurement of the electron density $n_e$. In the spectrum of {\tt Tr} the blue components of both doublets are absent, implying 3-sigma limits of 
$<0.07$ and $<0.04$ for 1883/1892 and 1907/1909 line ratios, respectively.
Such values correspond to $n_e \gtrsim 10^{6}$~cm$^{-3}$ \citep[e.g.,][]{maseda17}. We note that such high density is reminiscent of what measured in the Weigelt blobs nearby the LBV star of $\eta$~Car, where a mixture of dozens fluorescent and forbidden lines are detected \citep[][]{zet12}.

It is worth noting that \citet{joh06} explained the \siiiired\ emission observed in $\eta$~Car as due to the Resonance-Enhanced Two-Photon Ionisation (RETPI) mechanism, that is the combination of H~\lya\ (1216) and H~\lyb\ pumping. They also suggest that such a two-photon process naturally explains the absence of the blue component of the silicon doublet (\siiiiblue), as due to a gA value (Einstein transition probability) six order of magnitude smaller than \siiiired {. It is not clear if the same is happening on {\tt Tr}, for which the \siiiiblue\ is absent, too; further investigation is needed and a proper monitoring of the time variability of the line would be needed. Therefore the RETPI mechanisms for silicon remains a possibility.}

\begin{figure}
\centering
\includegraphics[width=8.5cm]{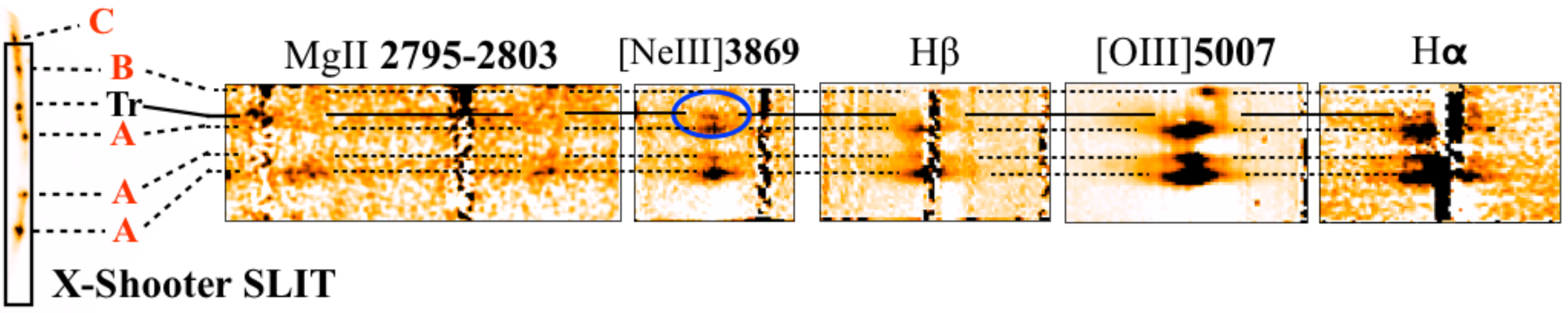}
\caption{Additional atomic transitions are shown, extracted from X-Shooter VIS and NIR arms. On the left side the $11$ arcsec X-Shooter slit is shown. For the knots A (the star cluster) and B, elements like Mg, Ne, H$\beta$, O and H$\alpha$ are detected in emission. For the transient ({\tt Tr}) only \neiii\ is clearly identified, despite the seeing limited X-Shooter observations, while the other lines are not identified. \neiii\ is marked with a blue ellipse. On the left the HST/F814W image with the X-Shooter slit are shown.}
\label{NIR}
\end{figure}

\section{Discussion}

From the above considerations it turns out that {\tt Tr} is likely a transient compact source, relatively bright with $-20.3<$~M$_{2000} < -18.6$ mag. Although the absolute magnitude of {\tt Tr} is not securely determined, it lies in a range which is typical of luminous SN explosions \citep{rich14}, while non-terminal eruptions of LBVs or other massive stars are expected to be fainter than $-15$ mag \citep{vand00,maun06}. Furthermore, the spectra of {\tt Tr} show fluorescent (\lya\ pumped) and forbidden \ferro\ emission lines in a relatively dense regime.
Interestingly, these fluorescence lines visible in the UV domain have been previously observed in some SNe with
evidence of strong interaction between the ejected gas and pre-existing circum-stellar material (CSM). Usually, these SNe
have spectra dominated by narrow (FWHM from a few tens to a few hundred \kms) H lines in emission, and are hence
labelled as Type IIn SNe. For instance, the UV lines observed in the spectra of {\tt Tr} (including \neiii\AA, see Figure~\ref{NIR}), were also
found in the HST UV spectra of the Type IIn SNe 1995N \citep[][]{fransson02} and 2010jl \citep[][]{fransson14}. However, it appears there is a deficiency of the Balmer lines \ha, \hb\ (Figure~\ref{NIR}).

The long-lasting light curve of {\tt Tr} (at least 1 year in rest-frame) is typical of SNe interacting with their CSM or
even long-duration giant eruptions of LBV stars. In addition, the FWHM of the fluorescence lines ($\simeq85$ \kms) is consistent
with the expansion velocity expected in the wind of an LBV. So, this would be an argument suggesting that {\tt Tr} is a long-duration eruption of a massive star, or even produced by a terminal stellar explosion in a hydrogen-rich circumstellar medium.
Although this explanation is not corroborated by a clear detection of Balmer lines in the {\tt Tr} spectrum (see Figure~\ref{NIR}), the detection of \heii\AA\ provides support to the CSM-interacting SN scenario. \heiialone\ lines in the optical domain are occasionally detected in young core-collapse SNe exploding inside a circum-stellar 
shell \citep{khaz16}.

The absence of the Balmer lines as well as \mgii\AA\ in the X-Shooter spectrum of {\tt Tr} is puzzling.
These lines are in fact expected in the spectrum of an H-rich  ejecta-CSM interacting transient
\citep[e.g.,][]{fransson05}. Alternatively, we are observing a lensed H-poor interacting SN.
SN Ibn \citep[e.g.,][]{pastore16} are well-known SNe interacting with a He-rich and H-deprived CSM whose spectra are dominated by relatively narrow and 
prominent \heialone\ features in emission. While these lines are not unequivocally
observed in the {\tt Tr} X-Shooter spectrum, \heii\AA\ is clearly detected in the June 2016 MUSE spectrum. In general, the \heiialone\ lines persist for a short period in SN spectra \citep[e.g.,][]{pastore15,gango20}, and this may explain their non-detection in 2019.
We also know that some ultrastripped core-collapse SNe may be powered by ejecta-CSM interaction without necessarily showing
H or \heialone\ lines. For instance, this scenario has been proposed for some super-luminous SNe \citep[e.g.,][]{moriya18} or in pulsational pair-instability events \citep[][]{woosley17}. However,
these transients should at least show prominent \oialonesq~6300,6364\AA, along with a blend \caiialonesq /\oii~at around 7300\AA.
In summary, the nature of {\tt Tr} is still unclear, 
although the presence of the object over the last year (in the rest-frame) and the similarity of the UV spectrum of {\tt Tr} with those of some interacting transients, most likely indicate a SN whose ejecta is interacting with some hydrogen-poor CSM.
The future monitoring of the light curve will be crucial to address the physical origin of this transient. 

Regardless of its true nature, {\tt Tr} demonstrates that the indirect detection of a gas condensation in the vicinity of a massive star at cosmological distance is feasible with gravitational lensing, even without the extremely high spatial resolution (i.e., of the order of a few light days), which would be required to observe it directly. 
Whether the gas condensation in {\tt Tr} is exposed to an external field with emission covering several Lyman series or it is invested by Lyman continuum that induces in-situ transitions within the condensation itself remains to be understood. 

In conclusion, {\tt Tr} is well recognised as an individual point-like object thanks to strong gravitational lensing, that is known to probe a limited volumes \citep[e.g.,][]{shu18}. However, future adaptive optics facilities (like E-ELT or VLT/MAVIS) will have enough angular resolution to detect even in blank fields objects which are currently observable only in lensed fields (with $\mu \sim 15-20$),
but embracing a much larger volume (a factor 10 larger then lensing) in the redshift range $1<z<8$, accessing absolute magnitudes down to  $\sim -17$.
This suggests that, potentially, cases like {\tt Tr} will be easily recognisable in the future.
 
\section*{Acknowledgments}
We thank the referee for the useful reports. This work is supported by PRIN-MIUR 2017 WSCC32.
%``Zooming into dark matter and proto-galaxies with massive lensing clusters''. 
We acknowledge funding from the INAF
%for ``interventi aggiuntivi a sostegno della ricerca di main-stream'' 
main-stream 2019. GBC acknowledges funding from the ERC Consolidator Grant ID 681627-BUILDUP.
We thank A. Comastri for useful discussions.
GB acknowledges funding for the Cosmic Dawn Center provided by the Danish National Research Foundation under grant No. 140.
HH acknowledges support from the Swedish research council VR under contract 2016-0418.

\section{Data availability} 
The data underlying this article will be shared on reasonable request to the corresponding author. They are also available as raw data at the European Southern Observatory archive under the programmes ID 0103.A-0688, 60.A-9507 and 297.A-5012, and HST archive under proposal ID 15101.


\begin{thebibliography}{}

%\bibitem[Annibali et al.(2015)]{annibali15} Annibali, F., Tosi, M., Pasquali, A., et al.\ 2015, \aj, 150, 143
%\bibitem[Alavi et al.(2016)]{alavi16} Alavi, A., Siana, B., Richard, J., et al.\ 2016, arXiv:1606.00469
%\bibitem[Atek et al.(2015)]{atek15} Atek, H., Richard, J., Jauzac, M., et al.\ 2015, \apj, 814, 69 
%\bibitem[Bacon et al.(2012)]{bacon12} Bacon, R., Accardo, M., Adjali, L., et al.\ 2012, The Messenger, 147, 4 
%\bibitem[Bacon et al.(2010)]{bacon10} Bacon, R., Accardo, M., Adjali, L., et al.\ 2010, \procspie, 7735, 773508
%\bibitem[Bacon et al.(2015)]{bacon15} Bacon R., et al., 2015, A\&A, 575, 75
%\bibitem[Bastian et al.(2013)]{bastian13} Bastian, N., Cabrera-Ziri, I., Davies, B., Larsen, S. S., 2013, MNRAS, 436, 2852
%\bibitem[Beckwith et al.(2006)]{beckwith06} Beckwith, S.~V.~W., Stiavelli, M., Koekemoer, A.~M., et al.\ 2006, \aj, 132, 1729 
\bibitem[Bowen(1934)]{bowen34} Bowen, I.~S.\ 1934, \pasp, 46, 146
%\bibitem[Bowen(1935)]{bowen35} Bowen, I.~S.\ 1935, \apj, 81, 1
\bibitem[Bowen(1947)]{bowen47} Bowen, I.~S.\ 1947, \pasp, 59, 196
%\bibitem[Alexandroff et al.(2013)]{alex13} Alexandroff, R., et al.\ 2013, \mnras, 435, 3306
%\bibitem[Behrens et al.(2014)]{behrens14} Behrens, C., Dijkstra, M., \& Niemeyer, J.~C.\ 2014, \aap, 563,77 
%\bibitem[Bekki \& Yong (2012)]{bekki12} Bekki, K., Yong, D., 2012, MNRAS, 419, 2063
%\bibitem[Bellini et al.(2015)]{bellini15} Bellini, A., Renzini, A., Anderson, J., et al.\ 2015, \apj, 805, 178
%\bibitem[Bertin et al.(2002)]{bertin02} Bertin, G., Ciotti, L., \& Del Principe, M.\ 2002, \aap, 386, 149 
%\bibitem[Bian et al.(2017)]{bian17} Bian, F., Fan, X., McGreer, I., Cai, Z., \& Jiang, L.\ 2017, \apjl, 837, L12 
%\bibitem[Borthakur et al.(2014)]{bort14} Borthakur, S., Heckman, T.~M., Leitherer, C., \& Overzier, R.~A.\ 2014, Science, 346, 216
%\bibitem[Bouwens et al.(2015)]{bouwens15} Bouwens, R. J., Illingworth, G. D., Oesch, P. A., Caruana, J., Holwerda, B., Smit, R., Wilkins, S., 811, 140
%\bibitem[Bouwens et al.(2016c)]{bouwens16c} Bouwens, R.~J., Oesch, P.~A., Illingworth, G.~D., Ellis, R.~S., \& Stefanon, M.\ 2016, arXiv:1610.00283 
%\bibitem[Bouwens et al.(2016b)]{bouwens16b} Bouwens, R.~J., Illingworth, G.~D., Oesch, P.~A., et al.\ 2016, arXiv:1608.00966  %% Extremely Small Sizes for Faint z~2-8 Galaxies in the HFF
%\bibitem[Bouwens et al.(2016)]{bouwens16} Bouwens, R., et al.\ 2016, \apj, 831, 176   %% The Lyman-Continuum Photon Production Efficiency
%\bibitem[Bragaglia et al.(2015)]{bragaglia15} Bragaglia, A., Carretta, E., Sollima, A., et al.\ 2015, \aap, 583, A69 
%\bibitem[Bridge et al.(2010)]{bridge10} Bridge, C.~R., Teplitz, H.~I., Siana, B., et al.\ 2010, \apj, 720, 465
%\bibitem[Brammer et al.(2016)]{brammer16} Brammer, G.~B., et al.\ 2016, \apjs, 226, 6 
%\bibitem[Bruzual \& Charlot(2003)]{BC03} Bruzual, G., \& Charlot, S.\ 2003, \mnras, 344, 1000 
%\bibitem[Calura et al.(2014)]{calu14} Calura, F., Ciotti, L., \& Nipoti, C.\ 2014, \mnras, 440, 3341 
%\bibitem[Calura et al.(2015)]{calu15} Calura, F., Few, C.~G., Romano, D., \& D'Ercole, A.\ 2015, \apjl, 814, L14
%\bibitem[Caminha et al.(2016a)]{cam16a} Caminha, G.~B., Grillo, C., Rosati, P., et al.\ 2016, \aap, 587, A80             %rxj2248   (a)
%\bibitem[Caminha et al.(2016b)]{cam16b} Caminha, G.~B., Karman, W., Rosati, P., et al.\ 2016, \aap, 595, A100   %nebula  (b)
%\bibitem[Caminha et al.(2016)]{caminha16} Caminha, G.~B., et al.\ 2016, \aap, 587, 80   %macs j0416 (c)
%\bibitem[Caminha et al.(2017)]{cam17} Caminha, G.~B., Grillo, C., Rosati, P., et al.\ 2017, \aap, 600, A90
%\bibitem[Castellano et al.(2012)]{castellano12} Castellano, M., Fontana, A., Grazian, A., et al.\ 2012, \aap, 540, A39 
%\bibitem[Castellano et al.(2014)]{castellano14} Castellano, M., Sommariva, V., Fontana, A., et al.\ 2014, \aap, 566, A19
%\bibitem[Caputi et al.(2017)]{caputi17} Caputi, K.~I., et al.\ 2017, \apj, 849, 45
%\bibitem[Castellano et al.(2016)]{castellano16} Castellano, M., et al.\ 2016b, \aap, 590, A31          %% ASTRODEEP
%\bibitem[Cen \& Kimm(2015)]{cen15} Cen, R., \& Kimm, T.\ 2015, \apjl, 801, L25
%\bibitem[Chevallard et al.(2017)]{chevallard17} Chevallard, J., et al.\ 2017, arXiv:1709.03503 
%\bibitem[Chisholm et al.(2017)]{chisholm17} Chisholm, J., Orlitov{\'a}, I., Schaerer, D., et al.\ 2017, \aap, 605,67 
%\bibitem[Chisholm et al.(2019)]{JC19} Chisholm, J., Rigby, J.~R., Bayliss, M., et al.\ 2019, \apj, 882, 182
%\bibitem[Castellano et al.(2016a)]{castellano16a} Castellano, M., Dayal, P., Pentericci, L., et al.\ 2016a, \apjl, 818, L3    %% reionization ionized bubbles
%\bibitem[Charbonnel et al.(2014)]{charbonnel14}Charbonnel, C., Chantereau, W., Krause, M., Primas, F., \& Wang, Y.\ 2014, \aap, 569, L6 
%\bibitem[Ciotti(1991)]{ciotti91} Ciotti, L.\ 1991, \aap, 249, 99 
%\bibitem[Ciotti(1994)]{ciotti94} Ciotti, L.\ 1994, Celestial Mechanics and Dynamical Astronomy, 60, 401 
%\bibitem[Cottrell \& Da Costa(1981)]{cottrell81} Cottrell, P.~L., \& Da Costa, G.~S.\ 1981, \apjl, 245, L79 
%\bibitem[Cristiani et al.(2016)]{cristiani16} Cristiani, S., Serrano, L.~M., Fontanot, F., Vanzella, E., \& Monaco, P.\ 2016, \mnras, 462, 2478 
\bibitem[Cupani et al.(2019)]{cupani19} Cupani, G., D'Odorico, V., Cristiani, S., et al.\ 2019, Astronomical Data Analysis Software and Systems XXVI, 362
%\bibitem[Cupani et al.(2015)]{cupani15} Cupani, G., D'Odorico, V., Cristiani, S., et al.\ 2015, Astronomical Data Analysis Software an Systems XXIV (ADASS XXIV), 289
\bibitem[Crowther(2007)]{crowther07} Crowther, P.~A.\ 2007, \araa, 45, 177
%\bibitem[Dawson et al.(2004)]{dawson04} Dawson, S., Rhoads, J.~E., Malhotra, S., et al.\ 2004, \apj, 617, 707
\bibitem[Dahle et al.(2016)]{dahle16} Dahle, H., Aghanim, N., Guennou, L., et al.\ 2016, \aap, 590, L4
%\bibitem[D'Antona \& Caloi(2004)]{dantona04} D'Antona, F., \& Caloi, V.\ 2004, \apj, 611, 871 
%\bibitem[D'Antona \& Caloi(2008)]{dantona08} D'Antona, F., \& Caloi, V.\ 2008, \mnras, 390, 693 
%\bibitem[Dayal \& Libeskind(2012)]{dayal12} Dayal, P., \& Libeskind, N.~I.\ 2012, \mnras, 419, L9 
%\bibitem[Dayal, \& Ferrara(2018)]{pratika18} Dayal, P., \& Ferrara, A.\ 2018, \physrep, 780, 1
%\bibitem[de Barros et al.(2016)]{debarros16} de Barros, S., et al.\ 2016, \aap, 585, A51 %Vanzella, E., Amor{\'{\i}}n, R., et al.\ 2016, \aap, 585, A51
%\bibitem[Decressin et al.(2007)]{decressin07} Decressin, T., Meynet, G., Charbonnel, C., Prantzos, N., \& Ekstr{\"o}m, S.\ 2007, \aap, 464, 1029 
%\bibitem[Decressin et al.(2010)]{decressin10} Decressin, T., Baumgardt, H., Charbonnel, C., \& Kroupa, P.\ 2010, \aap, 516, A73 
%\bibitem[Dijkstra, \& Wyithe(2007)]{dijkstra07} Dijkstra, M., \& Wyithe, J.~S.~B.\ 2007, \mnras, 379, 1589
%\bibitem[Dijkstra et al.(2016)]{dijkstra16} Dijkstra, M., Gronke, M., \& Venkatesan, A.\ 2016, \apj, 828, 71 
%\bibitem[de Mink et al.(2009)]{demink09} de Mink, S.~E., Pols, O.~R., Langer, N., \& Izzard, R.~G.\ 2009, \aap, 507, L1 
%\bibitem[Denissenkov \& Hartwick(2014)]{denissenkov14} Denissenkov, P.~A., \& Hartwick, F.~D.~A.\ 2014, \mnras, 437, L21 
%\bibitem[D'Ercole et al.(2008)]{dercole08} D'Ercole, A., Vesperini, E., D'Antona, F., McMillan, S.~L.~W., \& Recchi, S.\ 2008, \mnras, 391, 825 
%\bibitem[D'Ercole et al.(2016)]{dercole16} D'Ercole, A., D'Antona, F., \& Vesperini, E.\ 2016, \mnras, 461, 4088 
%\bibitem[Downing \& Sills(2007)]{downing07} Downing, J.~M.~B., \& Sills, A.\ 2007, \apj, 662, 341 
%\bibitem[Ellis et al.(2001)]{ellis01} Ellis, R., Santos, M.~R., Kneib, J.-P., \& Kuijken, K.\ 2001, \apjl, 560, L119 
%\bibitem[Erb et al.(2014)]{erb14} Erb, D.~K., Steidel, C.~C., Trainor, R.~F., et al.\ 2014, \apj, 795, 33
%\bibitem[Fall \& Rees(1985)]{fall85} Fall, S.~M., \& Rees, M.~J.\ 1985, \apj, 298, 18 
%\bibitem[Ferrara \& Loeb(2013)]{ferrara13} Ferrara, A., \& Loeb, A.\ 2013, \mnras, 431, 2826
%\bibitem[Finlator et al.(2017)]{finlator17} Finlator, K., Prescott, M.~K.~M., Oppenheimer, B.~D., et al.\ 2017, \mnras, 464, 1633
%\bibitem[Forte et al.(2012)]{forte12} Forte, J.~C., Vega, E.~I., \& Faifer, F.\ 2012, \mnras, 421, 635 
\bibitem[Fransson et al.(2002)]{fransson02} Fransson, et al.\ 2002, \apj, 572, 350
\bibitem[Fransson et al.(2014)]{fransson14} Fransson, C., Ergon, M., Challis, P.~J., et al.\ 2014, \apj, 797, 118
\bibitem[Fransson et al.(2005)]{fransson05} Fransson, C., et al.\ 2005, \apj, 622, 991
%\bibitem[Frohmaier et al.(2019)]{froh19} Frohmaier, C., Sullivan, M., Nugent, P.~E., et al.\ 2019, \mnras, 486, 2308
%\bibitem[Giallongo et al.(2015)]{giallongo15} Giallongo, E., Grazian, A., Fiore, F., et al.\ 2015, \aap, 578, A83
%\bibitem[Gal-Yam, A. et al.(2014)]{galy14} Gal-Yam, A., et al.\ 2014, Nature, 509, 471  
\bibitem[Gangopadhyay et al.(2020)]{gango20} Gangopadhyay, A., et al.\ 2020, \apj, 889, 170
\bibitem[Graham et al.(2017)]{gra17} Graham, M.~L., et al.\ 2017, \mnras, 469, 1559
%\bibitem[Gratton et al.(2004)]{gratton04} Gratton, R., Sneden, C., \& Carretta, E.\ 2004, \araa, 42, 385 
%\bibitem[Grazian et al.(2016)]{grazian16} Grazian, A., Giallongo, E., Gerbasi, R., et al.\ 2016, \aap, 585, A48
%\bibitem[Grazian et al.(2017)]{grazian17} Grazian, A., et al.\ 2017, \aap, 602, A18. %Giallongo, E., Paris, D., et al.\ 2017, \aap, 602, A18
\bibitem[Grillo et al.(2016)]{grillo16} Grillo, C., Karman, W., Suyu, S.~H., et al.\ 2016, \apj, 822, 78
\bibitem[Hamann et al.(1999)]{H99} Hamann, F., Davidson, K., Ishibashi, K., et al.\ 1999, Eta Carinae at the Millennium, 116
%\bibitem[Hashimoto et al.(2017)]{hashimoto17} Hashimoto, T., Ouchi, M., Shimasaku, K., et al.\ 2017, \mnras, 465, 1543
%\bibitem[Heckman et al.(2011)]{h11} Heckman, T.~M., et al.\ 2011, \apj, 730, 5.    %Borthakur, S., Overzier, R., et al.\ 2011, \apj, 730, 5
%\bibitem[Henry et al.(2015)]{henry15} Henry, A., Scarlata, C., Martin, C.~L., \& Erb,D.\ 2015, \apj, 809,19 
%\bibitem[Herenz et al.(2017)]{herenz17} Herenz, E.~C., Hayes, M., Papaderos, P., et al.\ 2017, \aap, 606, L11
%\bibitem[Humphrey et al.(2019)]{humphrey19} Humphrey, A., Villar-Mart{\'\i}n, M., Binette, L., et al.\ 2019, \aap, 621, A10
%\bibitem[Inoue et al.(2014)]{inoue14} Inoue, A.~K., Shimizu, I., Iwata, I., \& Tanaka, M.\ 2014, \mnras, 442, 1805 
%\bibitem[Inoue(2011)]{inoue11} Inoue, A.~K.\ 2011, \mnras, 415, 2920
%\bibitem[Izotov et al.(2016a)]{izotov16a} Izotov, Y.~I., Schaerer, D., Thuan, T.~X., et al.\ 2016a, \mnras, 461, 3683 
%\bibitem[Izotov et al.(2016b)]{izotov16b} Izotov, Y.~I., Orlitov{\'a}, I., Schaerer, D., et al.\ 2016b, \nat, 529, 178 
%\bibitem[Izotov et al.(2017)]{izotov17} Izotov, Y.~I., et al.\ 2017, arXiv:1711.11449
%\bibitem[Jaskot \& Oey(2013)]{jaskot13} Jaskot, A.~E., \& Oey, M.~S.\ 2013, \apj, 766, 91 
%\bibitem[Jaskot \& Oey(2014)]{jaskot14} Jaskot, A.~E., \& Oey, M.~S.\ 2014, \apjl, 791, L19
%\bibitem[Jaskot et al.(2017)]{jaskot17} Jaskot, A.~E., Oey, M.~S., Scarlata, C., \& Dowd, T.\ 2017, arXiv:1711.09516
%\bibitem[Laursen et al.(2011)]{laursen11} Laursen, P., Sommer-Larsen, J., \& Razoumov, A.~O.\ 2011, \apj, 728, 52
%\bibitem[Jimenez, \& Haiman(2006)]{JH06} Jimenez, R., \& Haiman, Z.\ 2006, \nat, 440, 501
\bibitem[Johansson et al.(2000)]{joh00} Johansson, S., Zethson, T., Hartman, H., et al.\ 2000, \aap, 361, 977
\bibitem[Johansson et al.(2006)]{joh06} Johansson, S., Hartman, H., \& Letokhov, V.~S.\ 2006, \aap, 452, 253
\bibitem[Johansson, \& Letokhov(2007)]{joh07} Johansson, S., \& Letokhov, V.~S.\ 2007, \nar, 51, 443
%\bibitem[Jones et al.(2013)]{jones13} Jones, T.~A., Ellis, R.~S., Schenker, M.~A., \& Stark, D.~P.\ 2013, \apj, 779, 52 
%\bibitem[Johnson(2010)]{J10} Johnson, J.~L.\ 2010, \mnras, 404, 1425
%\bibitem[Karman et al.(2015)]{karman15} Karman, W., Caputi, K.~I., Grillo, C., et al.\ 2015, \aap, 574, A11 
%\bibitem[Karman et al.(2016)]{karman16} Karman, W., Caputi, K.~I., Caminha, G.~B., et al.\ 2016, A\&A, accepted, arXiv:1606.01471 
%\bibitem[Kashikawa et al.(2012)]{kashi12} Kashikawa, N., Nagao, T., Toshikawa, J., et al.\ 2012, \apj, 761, 85
%\bibitem[Kaurov et al.(2019)]{kaurov19} Kaurov, A.~A., Dai, L., Venumadhav, T., et al.\ 2019, \apj, 880, 58
%\bibitem[Katz \& Ricotti(2014)]{katz14} Katz, H., \& Ricotti, M.\ 2014, \mnras, 444, 2377 
%\bibitem[Kawamata et al.(2015)]{kawamata15} Kawamata, R., Ishigaki, M., Shimasaku, K., Oguri, M., \& Ouchi, M.\ 2015, \apj, 804, 103 
\bibitem[Kelly et al.(2015)]{kelly15} Kelly, P.~L., et al.\ 2015, Science, 347, 1123
%\bibitem[Kelly et al.(2018)]{kelly18} Kelly, P.~L., Diego, J.~M., Rodney, S., et al.\ 2018, Nature Astronomy, 2, 334
%\bibitem[Chen et al.(2019)]{chen19} Chen, W., Kelly, P.~L., Diego, J.~M., et al.\ 2019, arXiv e-prints, arXiv:1902.05510
\bibitem[Khazov et al.(2016)]{khaz16} Khazov, D., et al.\ 2016, \apj, 818, 3 
%\bibitem[Kimm \& Cen(2014)]{kimm14} Kimm, T., \& Cen, R.\ 2014, \apj, 788, 121
%\bibitem[King(1962)]{king62} King, I.\ 1962, \aj, 67, 471 
%\bibitem[Koekemoer et al.(2014)]{koekemoer14} Koekemoer, A.~M., Avila, R.~J., Hammer, D., et al.\ 2014, American Astronomical Society Meeting Abstracts \#223, 223, 254.02 
%\bibitem[\protect\citeauthoryear{Kneib \& Natarajan}{2011}]{kneib11} Kneib J.-P., Natarajan P., 2011, A\&ARv, 19, 47 
%\bibitem[Krause et al.(2012)]{krause12} Krause, M., Charbonnel, C., Decressin, T., et al.\ 2012, \aap, 546, L5 
%\bibitem[Kravtsov \& Gnedin(2005)]{kravtsov05} Kravtsov, A.~V., \& Gnedin, O.~Y.\ 2005, \apj, 623, 650  %% GCs simulations
%\bibitem[Kroupa(2001)]{kroupa01} Kroupa, P.\ 2001, \mnras, 322, 231 
%\bibitem[Kruijssen(2015)]{kruijssen15} Kruijssen, J.~M.~D.\ 2015, \mnras, 454, 1658 
\bibitem[Leloudas et al.(2019)]{lelo19} Leloudas, G., et al.\ 2019, \apj, 887, 218
%\bibitem[Landt et al.(2008)]{landt08} Landt, H., Bentz, M.~C., Ward, M.~J., et al.\ 2008, \apjs, 174, 282 %Syfert bowen effect
%\bibitem[Lardo et al.(2011)]{lardo11} Lardo, C., Bellazzini, M., Pancino, E., et al.\ 2011, \aap, 525, A114 
%\bibitem[Larson(1969)]{larson69} Larson, R.~B.\ 1969, \mnras, 145, 405 
%\bibitem[Leitherer et al.(2014)]{leitherer14} Leitherer, C., Ekstr{\"o}m, S., Meynet, G., et al.\ 2014, \apjs, 212, 14 
%\bibitem[Leitherer et al.(2016)]{leitherer16} Leitherer, C., Hernandez, S., Lee, J.~C., \& Oey, M.~S.\ 2016, \apj, 823, 64
%\bibitem[Leitet et al.(2013)]{leitet13} Leitet, E., Bergvall, N., Hayes, M., Linn{\'e}, S., \& Zackrisson, E.\ 2013, \aap, 553, A106 
%\bibitem[Livermore et al.(2015)]{livermore15} Livermore, R.~C., Jones, T.~A., Richard, J., et al.\ 2015, \mnras, 450, 1812 
%\bibitem[Livermore et al.(2016)]{livermore16} Livermore, R.~C., Finkelstein, S.~L., \& Lotz, J.~M.\ 2016, arXiv:1604.06799
%\bibitem[Lotz et al.(2014)]{lotz14} Lotz, J., Mountain, M., Grogin, N.~A., et al.\ 2014, American Astronomical Society Meeting Abstracts \#223, 223, 254.01 
%\bibitem[Lotz et al.(2017)]{lotz17} Lotz, J.~M., et al.\ 2017, \apj, 837, 97   %Koekemoer, A., Coe, D., et al.\ 2017, \apj, 837, 97
\bibitem[Marziani et al.(2014)]{marziani14} Marziani, P., Mart{\'\i}nez-Aldama, M.~L., Dultzin, D., Sulentic, J.~W.\ 2014, The Astronomical Review, 9, 29
\bibitem[Maseda et al.(2017)]{maseda17} Maseda, M.~V., Brinchmann, J., Franx, M., et al.\ 2017, \aap, 608, A4
\bibitem[McClintock et al.(1975)]{clinto75} McClintock, J.~E., Canizares, C.~R., \& Tarter, C.~B.\ 1975, \apj, 198, 641
\bibitem[Merrill(1956)]{merrill56} Merrill, P.~W.\ 1956, \jrasc, 50, 184
\bibitem[\protect\citeauthoryear{Modigliani, et al.}{2010}]{modigliani10} Modigliani A., et al., 2010, SPIE, 7737, 773728, SPIE.7737
\bibitem[Moriya et al.(2018)]{moriya18} Moriya, T.~J., Sorokina, E.~I., \& Chevalier, R.~A.\ 2018, \ssr, 214, 59
%\bibitem[Madau \& Haardt(2015)]{madau15} Madau, P., \& Haardt, F.\ 2015, \apjl, 813, L8 
%\bibitem[Malhotra, \& Rhoads(2002)]{mal02} Malhotra, S., \& Rhoads, J.~E.\ 2002, \apjl, 565, L71
%\bibitem[Maseda et al.(2013)]{maseda13} Maseda, M.~V., van der Wel, A., da Cunha, E., et al.\ 2013, \apjl, 778, L22 
%\bibitem[Mas-Ribas et al.(2016)]{masribas16} Mas-Ribas, L., Dijkstra, M., \& Forero-Romero, J.~E.\ 2016, \apj, 833, 65
%\bibitem[Maseda et al.(2014)]{maseda14} Maseda, M.~V., van der Wel, A., Rix, H.-W., et al.\ 2014, \apj, 791, 17
%\bibitem[Mashchenko \& Sills(2005)]{mashchenko05} Mashchenko, S., \& Sills, A.\ 2005, \apj, 619, 243 
\bibitem[Maund et al.(2006)]{maun06} Maund, J.~R., et al.\ 2006, \mnras, 369, 390
%\bibitem[\protect\citeauthoryear{Meneghetti et al.}{2008}]{Meneghetti08} Meneghetti M., et al., 2008, A\&A, 482, 403 
%\bibitem[\protect\citeauthoryear{Meneghetti et al.}{2010}]{Meneghetti10} Meneghetti M., Rasia E., Merten J., Bellagamba F., Ettori S., Mazzotta P., Dolag K., Marri S., 2010, A\&A, 514, A93 
%\bibitem[Meneghetti et al.(2016)]{meneghetti16} Meneghetti, M., Natarajan, P., Coe, D., et al.\ 2016, arXiv:1606.04548 
%\bibitem[Merlin et al.(2016)]{merlin16} Merlin, E., Amor{\'{\i}}n, R., Castellano, M., et al.\ 2016, \aap, 590, A30 
%\bibitem[Micheva et al.(2017)]{micheva17} Micheva, G., Oey, M.~S., Jaskot, A.~E., \& James, B.~L.\ 2017, \apj, 845, 165
%\bibitem[Nagao et al.(2008)]{nagao08} Nagao, T., Sasaki, S.~S., Maiolino, R., et al.\ 2008, \apj, 680, 100
%\bibitem[Nakajima \& Ouchi(2014)]{nakajima14} Nakajima, K., \& Ouchi, M.\ 2014, \mnras, 442, 900 
%\bibitem[Nakasato et al.(2000)]{nakasato00} Nakasato, N., Mori, M., \& Nomoto, K.\ 2000, \apj, 535, 776 
\bibitem[Netzer et al.(1985)]{netzer85} Netzer, H., Elitzur, M., \& Ferland, G.~J.\ 1985, \apj, 299, 752
\bibitem[Pastorello et al.(2002)]{pastore02} Pastorello, A., et al.\ 2002, \mnras, 333, 27
\bibitem[Pastorello et al.(2015)]{pastore15} Pastorello, A., et al.\ 2015, \mnras, 449, 1941
\bibitem[Pastorello et al.(2016)]{pastore16} Pastorello, A., et al.\ 2016, \mnras, 456, 853
%\bibitem[Pacucci et al.(2017)]{pacucci17} Pacucci, F., Pallottini, A., Ferrara, A., et al.\ 2017, \mnras, 468, L77
%\bibitem[Pallottini et al.(2015)]{pallottini15} Pallottini, A., Ferrara, A., Pacucci, F., et al.\ 2015, \mnras, 453, 2465
%\bibitem[Peng et al.(2002)]{peng02} Peng, C.~Y., Ho, L.~C., Impey, C.~D., \& Rix, H.-W.\ 2002, \aj, 124, 266
%\bibitem[Peng et al.(2010)]{peng10} Peng, C.~Y., Ho, L.~C., Impey, C.~D., \& Rix, H.-W.\ 2010, \aj, 139, 2097 
%\bibitem[Piotto et al.(2007)]{piotto07} Piotto, G., Bedin, L.~R., Anderson, J., et al.\ 2007, \apjl, 661, L53 
%\bibitem[Plazas et al.(2019)]{pizas19} Plazas, A.~A., Meneghetti, M., Maturi, M., et al.\ 2019, \mnras, 482, 2823
%\bibitem[Prantzos \& Charbonnel(2006)]{prantzos06} Prantzos, N., \& Charbonnel, C.\ 2006, \aap, 458, 135
%\bibitem[Raiter et al.(2010)]{raiter10} Raiter, A., Schaerer, D., \& Fosbury, R.~A.~E.\ 2010, \aap, 523, A64
%\bibitem[Renzini et al.(2015)]{renzini15} Renzini, A., D'Antona, F., Cassisi, S., et al.\ 2015, \mnras, 454, 4197
%\bibitem[Rhoads et al.(2014)]{rhoads14} Rhoads, J.~E., Malhotra, S., Richardson, M.~L.~A., et al.\ 2014, \apj, 780, 20
%\bibitem[Rodney et al.(2018)]{rodney18} Rodney, S.~A., Balestra, I., Bradac, M., et al.\ 2018, Nature Astronomy, 2, 324
%\bibitem[Ricotti(2002)]{ricotti02} Ricotti, M.\ 2002, \mnras, 336, L33 
%\bibitem[Ricotti et al.(2016)]{ricotti16} Ricotti, M., Parry, O.~H., \& Gnedin, N.~Y.\ 2016, \apj, 831, 204
%\bibitem[Rydberg et al.(2013)]{rydberg13} Rydberg, C.-E., Zackrisson, E., Lundqvist, P., et al.\ 2013, \mnras, 429, 3658
%\bibitem[Rydberg et al.(2015)]{rydberg15} Rydberg, C.-E., Zackrisson, E., Zitrin, A., et al.\ 2015, \apj, 804, 13. %% CLASH and PopIII
\bibitem[Richardson et al.(2014)]{rich14} Richardson, D., Jenkims, R.~L. III, Wright, J., Maddox, L.\ 2014, \aj, 147, 118 
%\bibitem[Rivera-Thorsen et al.(2017)]{rivera17} Rivera-Thorsen, T.~E., et al.\ 2017, arXiv:1710.09482 
\bibitem[Rivera-Thorsen et al.(2019)]{rivera19} Rivera-Thorsen, T.~E., et al.\ 2019, Science, 366, 738
\bibitem[Shu et al.(2018)]{shu18} Shu, Y., Bolton, A.~S., Mao, S., et al.\ 2018, \apj, 864, 91
\bibitem[Sosnowska et al.(2015)]{sosnowska15} Sosnowska, D., Lovis, C., et al.\ 2015, Astronomical Data Analysis Software an Systems XXIV (ADASS XXIV), 285
%\bibitem[Sosnowska et al. (2015)]{sosnowska15} Sosnowska, D., Lovis, C., Figueira, P., et al.\ 2015, \aspconf, 495, 285
% Dahle, H., Gronke, M., et al.\ 2017, arXiv:1710.09482   %% triple Lya
%\bibitem[Rivera-Thorsen et al.(2017a)]{rivera17a} Rivera-Thorsen, T.~E., {\"O}stlin, G., Hayes, M., \& Puschnig, J.\ 2017, \apj, 837, 29  %%Haro11, channel
%\bibitem[Robert et al.(2003)]{robert03} Robert, C., Pellerin, A., Aloisi, A., et al.\ 2003, \apjs, 144, 21
%\bibitem[Robertson et al.(2015)]{robertson15} Robertson, B.~E., Ellis, R.~S., Furlanetto, S.~R., \& Dunlop, J.~S.\ 2015, \apjl, 802, L19 
%\bibitem[Salpeter(1955)]{salpeter55} Salpeter, E.~E.\ 1955, \apj, 121, 161 
%\bibitem[Schaerer et al.(2016)]{schaerer16} Schaerer, D., Izotov, Y.~I., Verhamme, A., et al.\ 2016,\aap, 591,8
%\bibitem[Schaerer(2002)]{schaerer02} Schaerer, D.\ 2002, \aap, 382, 28 
%\bibitem[Schaerer(2003)]{schaerer03} Schaerer, D.\ 2003, \aap, 397, 527 
%\bibitem[Schaerer \& de Barros(2009)]{schaerer09} Schaerer, D., \& de Barros, S.\ 2009, \aap, 502, 423 
%\bibitem[Schaerer \& Charbonnel(2011)]{schaerer11} Schaerer, D., \& Charbonnel, C.\ 2011, \mnras, 413, 2297 
%\bibitem[Shapley et al.(2016)]{shapley16} Shapley, A., Steidel, C.~C., Strom, A.~L., et al.\ 2016, \apjl, 826,24
%\bibitem[Shibuya et al.(2018)]{shi18} Shibuya, T., Ouchi, M., Harikane, Y., et al.\ 2018, \pasj, 70, S15
%\bibitem[Shivaei et al.(2017)]{shivaei17} Shivaei, I., et al.\ 2017, arXiv:1711.00013 
%\bibitem[Siana et al.(2010)]{siana10} Siana, B., Teplitz, H.~I., Ferguson, H.~C., et al.\ 2010, \apj, 723, 241 
%\bibitem[Siana et al.(2015)]{siana15} Siana, B., et al.\ 2015, \apj, 804, 17. %Shapley, A.~E., Kulas, K.~R., et al.\ 2015, \apj, 804, 17
%\bibitem[Smit et al.(2015)]{smit15} Smit, R., Bouwens, R.~J., Franx, M., et al.\ 2015, \apj, 801, 122 
%\bibitem[Sobral et al.(2015)]{sobral15} Sobral, D., Matthee, J., Darvish, B., et al.\ 2015, \apj, 808, 139
%\bibitem[Sobral et al.(2019)]{sobral19} Sobral, D., Matthee, J., Brammer, G., et al.\ 2019, \mnras, 482, 2422
%\bibitem[Stark et al.(2015)]{stark15} Stark, D.~P., et al.\ 2015, \mnras, 454, 1393
%\bibitem[Stark et al.(2017)]{stark17} Stark, D.~P., et al.\ 2017, \mnras, 464, 469 
%\bibitem[Steidel et al.(2001)]{steidel01} Steidel, C.~C., Pettini, M., \& Adelberger, K.~L.\ 2001, \apj, 546, 665 
%\bibitem[Terlevich et al.(2016)]{terlevich16} Terlevich, R., Melnick, J., Terlevich, E., et al.\ 2016, \aap, 592, L7 
%\bibitem[Tornatore et al.(2007)]{torna07} Tornatore, L., Ferrara, A., \& Schneider, R.\ 2007, \mnras, 382, 945
%\bibitem[Trakhtenbrot et al.(2019)]{trak19} Trakhtenbrot, B., Arcavi, I., Ricci, C., et al.\ 2019, Nature Astronomy, 3, 242
%\bibitem[Trenti et al.(2015)]{trenti15} Trenti, M., Padoan, P., \& Jimenez, R.\ 2015, \apjl, 808, L35 
%\bibitem[Treu et al.(2015)]{treu15} Treu, T., Schmidt, K.~B., Brammer, G.~B., et al.\ 2015, \apj, 812, 114 
\bibitem[Van Dyk et al.(2000)]{vand00} Van Dyk, S.~D., et al.\ 2000, \pasp, 112, 1532 
%\bibitem[Vanzella et al.(2010)]{vanz10} Vanzella, E., Giavalisco, M., Inoue, A.~K., et al.\ 2010, \apj, 725, 1011   %% Ultradeep U-band fesc
%\bibitem[Vanzella et al.(2010)]{vanz10} Vanzella, E., Siana, B., Cristiani, S., \& Nonino, M.\ 2010, \mnras, 404, 1672 %% CONTMAINANTS
%\bibitem[Vanzella et al.(2012a)]{vanz12a} Vanzella, E., Nonino, M., Cristiani, S., et al.\ 2012, \mnras, 424, L54  %% fesc and strong lensing
%\bibitem[Vanzella et al.(2012)]{vanz12} Vanzella, E.,  et al.\ 2012, \apj, 751, 70   %Guo, Y., Giavalisco, M., et al.\ 2012, \apj, 751, 70  %% CANDELS ionizing sources
%\bibitem[Vanzella et al.(2014)]{vanz14} Vanzella, E.,  et al.\ 2014, \apjl, 783, L12  %Fontana, A., Zitrin, A., et al.\ 2014, \apjl, 783, L12   %% very faint SF galaxy at z=6.4, MAS J0717
%\bibitem[Vanzella et al.(2015)]{vanz15} Vanzella, E., et al.\ 2015, \aap, 576, A116 %de Barros, S., Castellano, M., et al.\ 2015, \aap, 576, A116  %% colors LBGs and Galfit fit
%\bibitem[Vanzella et al.(2016c)]{vanz16c} Vanzella, E., Balestra, I., Gronke, M., et al.\ 2016c, arXiv:1607.03112   %   Lya nebula
%\bibitem[Vanzella et al.(2016)]{vanz16} Vanzella, E., et al.\ 2016, \apj, 825, 41 %de Barros, S., Vasei, K., et al.\ 2016b, \apj, 825, 41            %   Io2, LyC
%\bibitem[Vanzella et al.(2016a)]{vanz16a} Vanzella, E., De Barros, S., Cupani, G., et al.\ 2016a, \apjl, 821, L27     %  ID11 X-Shooter
%\bibitem[Vanzella et al.(2017)]{vanz17} Vanzella, E., Calura, F., Meneghetti, M., et al.\ 2017, \mnras, 467, 4304   %% Proto-GC
%\bibitem[Vanzella et al.(2019)]{vanz19} Vanzella, E., Calura, F., Meneghetti, M., et al.\ 2019, \mnras, 483, 3618
\bibitem[Vanzella et al.(2020)]{vanz20} Vanzella, E., et al.\ 2020, \mnras, 491, 1093
%\bibitem[Vanzella et al.(2017)]{vanz17} Vanzella, E., et al.\ 2017, \apj, 842, 47   %%Castellano, M., Meneghetti, M., et al.\ 2017, \apj, 842, 47 %% ID14
%\bibitem[Vanzella et al.(2017)]{vanz17a} Vanzella, E., Balestra, I., Gronke, M., et al.\ 2017, \mnras, 465, 3803   %% Illuimnating, HUDF
%\bibitem[Verhamme et al.(2016)]{verhamme16} Verhamme, A., Orlitova, I., Schaerer, D., et al.\ 2016, arXiv:1609.03477
%\bibitem[Verhamme et al.(2015)]{verhamme15} Verhamme, A., Orlitov{\'a}, I., Schaerer, D., \& Hayes, M.\ 2015, \aap, 578, A7 
%\bibitem[Verhamme et al.(2017)]{verhamme17} Verhamme, A., Orlitov{\'a}, I., Schaerer, D., et al.\ 2017, \aap, 597,13 
%\bibitem[Vesperini \& Heggie(1997)]{vesperini97} Vesperini, E., \& Heggie, D.~C.\ 1997, \mnras, 289, 898 
\bibitem[Wallerstein et al.(1991)]{walle91} Wallerstein, G., Schachter, J., Garnavich, P.~M., Oke, J. B.\ 1991, \pasp, 103, 185
%\bibitem[Wang et al.(2004)]{wang04} Wang, W., Liu, X.-W., Zhang, Y., et al.\ 2004, \aap, 427, 873
\bibitem[Weigelt, \& Ebersberger(1986)]{w86} Weigelt, G., \& Ebersberger, J.\ 1986, \aap, 163, L5
%\bibitem[Wise et al.(2012)]{wise12} Wise, J.~H., Turk, M.~J., Norman, M.~L., et al.\ 2012, \apj, 745, 50
%\bibitem[Wise et al.(2014)]{wise14} Wise, J.~H., Demchenko, V.~G., Halicek, M.~T., et al.\ 2014, \mnras, 442, 2560
%\bibitem[Wise(2019)]{wise19} Wise, J.~H.\ 2019, arXiv e-prints, arXiv:1907.06653
%\bibitem[Wolf et al.(2010)]{wolf10} Wolf, J., Martinez, G.~D., Bullock, J.~S., et al.\ 2010, \mnras, 406, 1220 
%\bibitem[Yue et al.(2014)]{yue14} Yue, B., Ferrara, A., Vanzella, E., \& Salvaterra, R.\ 2014, \mnras, 443, L20
%\bibitem[Worseck et al.(2014)]{worseck14} Worseck, G.,  et al.\ 2014, \mnras, 445, 1745. %%Prochaska, J.~X., O'Meara, J.~M., et al.\ 2014, \mnras, 445, 1745
%\bibitem[Yang et al.(2016)]{yang16} Yang, H., Malhotra, S., Gronke, M., et al.\ 2016, \apj, 820, 130
\bibitem[Weymann, \& Williams(1969)]{wey69} Weymann, R.~J., \& Williams, R.~E.\ 1969, \apj, 157, 1201
\bibitem[Woosley(2017)]{woosley17} Woosley, S.~E.\ 2017, \apj, 836, 244
%\bibitem[Windhorst et al.(2018)]{wind18} Windhorst, R.~A., Timmes, F.~X., Wyithe, J.~S.~B., et al.\ 2018, \apjs, 234, 41
%\bibitem[Zackrisson et al.(2011)]{zack11} Zackrisson, E., Inoue, A.~K., Rydberg, C.-E., et al.\ 2011, \mnras, 418, L104
%\bibitem[Zackrisson et al.(2013)]{zack13} Zackrisson, E., Inoue, A.~K., \& Jensen, H.\ 2013, \apj, 777, 39
%\bibitem[Zackrisson et al.(2015)]{zack15} Zackrisson, E., Gonz{\'a}lez, J., Eriksson, S., et al.\ 2015, \mnras, 449, 3057  %% star clusters magnified PopIII
%\bibitem[Zackrisson, \& Vikaeus(2019)]{zack19} Zackrisson, E., \& Vikaeus, A.\ 2019, arXiv e-prints, arXiv:1903.12555
%\bibitem[Zackrisson et al.(2017)]{zackrisson17} Zackrisson, et al.\ 2017, \apj, 836, 78   %%E., Binggeli, C., Finlator, K., et al.\ 2017, \apj, 836, 78   %% LyC leakage - Hbeta - uv slope connetion
\bibitem[Zethson et al.(2012)]{zet12} Zethson, T., Johansson, S., Hartman, H., Gull, T. R.\ 2012, \aap, 540, A133
\end{thebibliography}
\end{document}